\documentclass[twocolumn,showpacs,preprintnumbers,amsmath,amssymb]{revtex4}

\usepackage{graphicx}
\usepackage{dcolumn}
\usepackage{bm}

\begin{document}

\title{Coherent manipulation of atoms by co-propagating laser beams}

\author{Yuri B. Ovchinnikov}
\affiliation{National Physical Laboratory, Hampton Road, Teddington, Middlesex, TW11 0LW, UK}

\email{yuri.ovchinnikov@npl.co.uk}

\date{\today}

\begin{abstract}
Optical dipole traps and fractional Talbot optical lattices based on the interference between multiple
co-propagating laser beams are proposed. The variation of relative amplitudes and phases of the
interfering light beams of these traps makes it possible to manipulate the spatial position of trapped atoms.
Examples of spatial translation and splitting of atoms between a set of the interference traps are considered.
The prospect of constructing all-light atom chips based on the proposed technique is presented.
\end{abstract}

\maketitle

\section{Introduction}
There are three general types of conservative atom traps: magnetic, electrostatic and optical dipole traps.
This paper concentrates on a special kind of all-light atom chip traps. First we review the
progress in magnetic surface traps, which have been the basis for atom chips to date.

Magnetic atom chips \cite{Folman} are becoming more and more
popular experimental technique to produce and manipulate
Bose-Einstein condensed (BEC) coherent ensembles of ultra-cold
atoms \cite{BEC} . The main advantage of magnetic chips is their
compactness, relatively low energy consumption and low cost, which
is related to the proximity of atoms to the sources of magnetic
field. Most of the chips use current caring wires to create strong
non-uniform magnetic fields. Initially magnetic chips were
used only to trap \cite{Hansch} and guide \cite{Prentiss} cold
thermal atoms, now the more complicated versions of them are
used to get BEC \cite{Hansch_2} and even to implement
integrated atom interferometers \cite{Schumm}. In spite the fact
that the current consumed by each of the elements of the magnetic
chip is essentially smaller compared to the macroscopic magnetic
traps, it is far above the currents used in traditional integrated
electronic elements. Further complication of magnetic
atom chips will sooner or later have to face up to the problem of
removing dissipated heat. The further miniaturization of the
magnetic chip elements is restricted by the fluctuations of the
magnetic field caused by thermally excited currents in the metal
wires and some other effects [8], which rapidly increase with
decreasing distance between the trapped atoms and the surface of
the chip.

All-light atom chips involve manipulation of atomic motion with light forces only. To provide coherent manipulation
of atoms in these chips far-detuned dipole traps \cite{GWO} have to be used. There are multiple advantages of these chips over
the magnetic chips. The all-light chips are free of the problems of heat dissipation and the absorption of light in
their substrates is rather small. The Van-der-Waals interaction of atoms with a surface of a dielectric substrate becomes
important only at sub-micron distances \cite{Aspect}. Another advantage of the light traps is that they can confine atoms in all
magnetic energy states including the magnetically in-sensitive ones. This last property of the all-light chips makes them very
advantageous for precise measurements. Finally, the cold ensembles of molecules can be formed and stored in these traps
\cite{Grimm}.

The development of all-light atom chips lags behind that of magnetic chips.
One approach to the manipulation of atoms near the surface of a dielectric substrate is based on evanescent light waves
\cite{Dowling}. In fact the earliest proposal of a surface trap for neutral atoms was a light dipole trap based on a
superposition of two evanescent light waves of different frequencies \cite{Ovchinnikov}. The first observation of atoms
cooled in an evanescent light trap was made in 1997 \cite{Ovchinnikov_2} after the cooling mechanism of atoms in
evanescent waves became available \cite{Soeding}. The evanescent light wave traps are still under development and a
recent result is that BEC Cs atoms have been produced in such a trap \cite{Rychtarik}, this was not possible
in magnetic traps due to specific properties of collisions between cold Cs atoms.
Recently also the surface waveguides \cite{Prentiss_2} and two-dimensional surface optical lattices \cite{Christandl}
based on the bichromatic evanescent waves have been proposed.

Another approach to the surface light traps, which was developed in parallel to evanescent wave traps, is based on
standing light waves \cite{Mlynek}. Recently an integrated atom-optical circuit, which combines cooling of atoms in
a surface magneto-optical trap and their subsequent transportation along the waveguide formed by a standing light wave,
has been demonstrated \cite{Mlynek_2}.
Finally, in \cite{Ertmer_1, Ertmer_2} a microscopic array of independent dipole light traps formed by an array of
microlenses was realized.

These light and magnetic chips technologies complement each other nicely, as demonstrated by the new generation of combined
atom chips \cite{Wang, Eriksson}.

This paper introduces another approach to the construction of near-surface all-light atom optics
elements based on the interference between multiple co-propagating laser beams. It also demonstrates
how to coherently transfer and split atoms between arrays of such traps, which opens the prospects of building new
interferometers, quantum registers and other missing elements of all-light chips.

Following this introduction the second section considers basic multi-beam traps, and concentrates on
a two-beam trap, because it is the simplest configuration interference trap. In the third
section the basic manipulations of the spatial position of the trapped atoms are considered. The fourth section considers
optical lattices based on a fractional Talbot effect. In the outlook section the prospects of building all-light
atom chips are briefly discussed. Finally, in the conclusion the results of this investigation are summarized.

\section{Multi-beam interference dipole traps}

The simplest trap based on the interference of co-propagating laser beams is a two-beam dipole trap.
Although the two-beam trap is not the most ideal trap of this kind, it is the best to explain the general principle.
The trap is formed by the two Gaussian laser beams, which are co-propagating in z direction (see Fig.\,1).
The waists of the beams locate in the X0Y plane.
Our interest is the region A, where the interference of the two beams forms the fringes, which can be used for
three-dimensional conservative trapping of atoms by dipole light forces \cite{GWO}.
The amplitude of the light field can be written as

\begin{equation}
\begin{split}
E(x,y,z)=\sum_{j=1}^n\sqrt{\frac{2}{\pi}} \frac{1}{w^2(z)} \exp \left[-\frac{(x-x_j)^2+(y-y_j)^2}{w^2(z)}\right]\times\\
\exp\left[-i \left( kz-\tan^{-1}(z/z_R)+\frac{k((x-x_j)^2+(y-y_j)^2)}{2(z+z_R^2/z)} \right) \right],
\end{split}
\end{equation}

where for a two-beam trap $n=2$, $x_1=-d/2$, $x_2=d/2$,
$y_{1,2}=0$,  $d$ is the distance between the beams, $w(z)=w_0
\sqrt{1+(z/z_R)^2}$ is the radius of each of the two beams, $w_0$
is their waist radiuses, $k=2 \pi/\lambda$ is the wave vector and
$z_R=\pi w_0^2/\lambda$ is the Rayleigh range. The amplitude of
each of the two beams is normalized to unit power. We suppose here
that both beams are of the same linear polarization directed along
the y-axis. The spatial distribution of the light intensity
resulted from the interference between the two beams is described
by the formula

\begin{equation}
\begin{split}
I(x,y,z)=\frac{2}{\pi w^2(z)}\times \exp\left(-\frac{d^2+4(x^2+y^2)}{2 w^2(z)}\right)\times\\
\left[\exp\left(-\frac{2 d x}{w^2(z)}\right)+
\exp\left(\frac{2 d x}{w^2(z)}\right)+
2\cos\left(\frac{2 \pi d x z}{\lambda (z^2+z_R^2)}\right)\right],
\end{split}
\end{equation}

The interference between the two beams forms intensity fringes along the x-axis, which are described by the cosine term
inside the square brackets. These fringes are very well known from the Young's double slit interferometer.
The first exponential term of equation (2) shows that the intensity of the central interference fringes reaches
maximum at $z \neq 0$. For the central fringe, when $x=0$ and $y=0$, the maximum of intensity is reached at

\begin{equation}
z=z_{max}=z_R \sqrt{\frac{d^2}{2 w_0^2}-1}.
\end{equation}

When the waist radii, $w_0$ of the beams and the distance, $d$ between them are comparable to each other,
the interference pattern consists mostly of a single central fringe. Fig.\,2 shows the intensity
distribution of the two-beam interference pattern in the X0Z plane for $z>0$, $w_0=\lambda$ and $d=3 \lambda$.
One can see that the central interference fringe is partially separated from the waist regions of the two
source beams. The cross sections of the spatial intensity distribution of the central interference
fringe along the x, y and z coordinate axis are shown on Fig.\,3\,a,\,b,\,c. The maximum of the light intensity in
this fringe is reached at
$z=5.66\,\lambda$ in a full agreement with the formula (3). Fig.\,3\,d shows the intensity distribution along the line
$z=z_{max} \pm 2 x z_{max}/d$, which connects the centre of the fringe to the centres of the waists of the source
beams. These two directions define the potential depth of the corresponding red-detuned dipole trap formed by the central
interference fringe. It is expected that the escape of atoms from the trap due to their heating or tunnelling will happen
mostly along these directions.

According to the formula (2) the interference trap is scalable. If the sizes of the interfering beams $w_0$ and the
distance between them $d$ are increased proportionally by the same factor $a$, the transverse size of the trap along
the x and y axes also increases by factor $a$, while the longitudinal size along the z-axis increases by factor $a^2$.
This scalability law is exactly the same as for a single Gaussian light beam. The situation when the sizes of the light
beams stays fixed and only the distance between them is increased is very different. In that case there additional
intensity fringes along the x-axis of the interference trap become visible.

To increase the degree of localization of the central interference fringe, a larger number of the
interfering light beams must be used.
Let us skip the three-beam case and consider the trap, which consists of four equidistant co-propagating
light beams with waist radiuses of $w_0=\lambda$, which are located at the corners of a square with side
length $d=3\,\lambda$, as it is shown in Fig.\,4. The amplitude of the interference field produced by the interference between the four
beams is described by the formula (1), where the coordinates of the centres of the four Gaussian beams $(x_j,\,y_j)$
are (-d/2,\,d/2), (-d/2,\,-d/2), (d/2,\,d/2) and (d/2,\,-d/2).
The spatial distribution of the light intensity in the r0Z plane, which is crossing the square trap along
its diagonal r (see Fig.\,4), is shown in Fig.\,5. The maximum of the intensity in the central interference
fringe is localized at $z=8.885 \lambda$. Fig.\,6\,a,\,b,\,c,\,d shows the spatial intensity distribution along four different directions
around the maximum of the fringe. The last graph shows the intensity distribution along the line which joins the waist
centre of one of the source beams and the centre of the interference fringe. One can see that the separation of the fringe is
nearly complete.

To estimate the potential depth of the red-detuned trap formed by the interference of the four co-propagating laser beams
we will take an example of Rb$^{87}$ atom trapped by the coherent light of a Nd:YAG laser with wavelength of
$\lambda=1.064\, \mu$m. We will use a two-level model of the atom, the wavelength of which $\lambda_0=0.780\, \mu$m
corresponds to the strongest $5S^{1/2} \rightarrow 5P^{3/2}$ transition of Rb$^{87}$. Such a frequency
of the trapping light is chosen to minimize the probability of the spontaneous scattering of photons by an atom and
therefore to preserve the coherence of the light-atom interaction.
The dipole potential formed by the light can be estimated from the formula \cite{GWO}

\begin{equation}
U_{dip}(x,y,z)=-\frac{3 \pi c^2}{2 \omega_0^3}\left(\frac{\Gamma}{\omega_0-\omega}+\frac{\Gamma}{\omega_0+\omega}\right) I(x,y,z),
\end{equation}

where $\omega_0=2 \pi c/ \lambda_0$ is the frequency of the atomic transition, $\omega$ is the frequency of the light field,
$\Gamma$ is the linewidth of the transition and $I(x,y,z)$ is the spatial distribution of the light intensity.
We neglect here the fine structure of the transition between the ground 5S state of Rb$^{87}$ and its first 5P excited
state, which causes just a small correction to the magnitude of the potential (4) for the chosen frequency of the trapping
light. For the laser power 1.6\,mW in each beam of the four-beam trap this formula gives the total depth of the central interference fringe of 82\,$\mu$K.
The corresponding oscillation frequencies of the Rb atoms around the minimum of the trap are equal to $f_x=f_y=18.7$\,kHz
and $f_z=2.6$\,kHz, which gives the mean frequency of the trap of $f_0=16.6$\,kHz. The rate of spontaneous
scattering of the trapped atoms can be estimated from the formula \cite{GWO}

\begin{equation}
\begin{split}
\Gamma_{scat}(x,y,z)= -\frac{3 \pi^2 c^2}{2 \hbar \omega_0^3}
\left( \frac{\omega}{\omega_0}\right)^3\\
\left(\frac{\Gamma}{\omega_0-\omega}+\frac{\Gamma}{\omega_0+\omega}\right)^2
I(x,y,z).
\end{split}
\end{equation}

For the parameters of the light field given above the minimum time between the spontaneous scattering events
$1/\Gamma_{scat}$ near the bottom of the trap is about 1 second.
To get longer coherence times without scarifying the potential depth, CO$_2$ lasers with a wavelength
of 10\,$\mu$m can be used \cite{THG}.

Additional heating of the trapped atoms can be caused by technical noise of the laser beams.
Periodic change of the depth of the dipole trap causes transition of the atoms to the higher
vibrational states of the trap potential. Such parametric heating of the atoms in the multi-beam interference trap
due to fluctuations of the light intensity is expected to be the same as in other dipole traps and
is relatively small \cite{Savard, Alt} even for free-running industrial lasers.

The other source of parametric heating is a motion of the trapping
dipole potential caused by fluctuations of the relative phases of
the interfering laser beams. The corresponding transition rate of
trapped atoms from the ground to the first vibration state of the
trap is

\begin{equation}
R_{0 \rightarrow 1}=\frac{2 \pi^3 M f_x^3}{\hbar} S_x(f_x),
\end{equation}

where $f_x$ is the transverse frequency of the trap, $M$ is the
mass of atom and $S_x(f_x)$ is the spectral density of the
position fluctuations in the trap centre along the $x$ axis
\cite{Gehm}. According to this equation, if an atom is confined in
a trap with an oscillation frequency $f_x=18.7\,$kHz, achievement
of the transition rate of $R_{0 \rightarrow 1}=1\,s^{-1}$ requires
a position stability of
$\sqrt{S_x(f_x)}=2.4\times10^{-6}\,\mu$m/$\sqrt{\textnormal{Hz}}$.
The change of the position of the interference trap along the $x$
axis $\epsilon=\alpha \Delta\phi$ is a linear function of the
phase difference between the interfering laser beams $\Delta\phi$,
where the proportionality coefficient $\alpha=D/2\pi$ is
determined by the period of the interference fringes $D \approx
z_{max} \lambda/d$ along the $x$ axis at $z=z_{max}$. For the
given above example of a four beam interference trap the period of
the interference fringes along $x$ axis at $z=z_{max}=9.45\,\mu$m
is $D \approx 3\,\mu$m. Therefore, to achieve $R_{0 \rightarrow
1}=1\,s^{-1}$, the required relative phase stability of the laser
beams is
$\sqrt{S_{\phi}(f_x)}=5\times10^{-6}\,\textnormal{rad}/\sqrt{\textnormal{Hz}}$.
According to \cite{Ye} the phase spectral density of a phase
locked Nd:YAG laser at frequency $f=1$\,kHz is better than
$\sqrt{S_{\phi}(f)}=3\times10^{-6}\,\textnormal{rad}/\sqrt{\textnormal{Hz}}$.
It means that even independent lasers can provide the
phase stability required. If all the trapping beams are generated from the
same laser, their relative phase fluctuations should be
essentially smaller than the absolute phase noise of the laser
light. In fact in standard optical lattices the demands on the
phase stability of the laser beams are stricter than in the
multibeam copropagating interference dipole traps for two
reasons. First, the typical vibration frequency of the optical
lattice potential is higher ($\ge$100\,kHz) and
according to the equation (6) the parametric heating rate of atoms
grows rapidly with the frequency. Second, all the beams of the
copropagating interference trap can be formed from different parts
of a single laser beam. It means the relative phases of these
beams will be as stable as the phase across a single laser
beam, which is not the case of standard optical lattices, where
the laser beams have to propagate along different passes before they
interfere with each other.

\section{Manipulation of atoms}

A coherent dipole light trap is a very useful element of atom
optics. However many important applications such as atom
interferometry \cite{Berman} or quantum computing \cite{QC}
require the ability to manipulate atoms between or within these
traps. In this paper only the manipulation of a spatial position
of atoms between the interference traps will be described. There
are two basic types of such manipulation, the translation of atoms
and their splitting between two or more dipole traps. Let us
consider a spatial translation of atoms first. Certainly the
simplest way to translate trapped atoms in space is by adiabatic
change of the spatial position of the trapping light beams, but
that is not possible for all-light atom chips, in which the
position of the light beams emitted from the surface of the chip
is fixed and determined by the design of the chip. The only
parameters, which can be changed are the intensities, phases,
polarizations or frequencies of the individual light beams.
Multi-beam interference traps offer the possibility of translating
atoms in all directions, but only the translation of atoms across
the light beams is considered here.

One way to move atoms across the interfering light beams is to
change their relative phases. For a two-beam interference trap the
change of the phase of one of the beams will lead to the
transverse displacement of the interference fringes. The problem
of the two-beam trap is that the shifted central interference
fringe eventually becomes merged with one of the source beams,
which leads to transfer of the trapped atoms into the waist region
of this beam. For an atom chip, where the waists of the source
beams are located in the plane of the surface of the chip, that
means atom reaching the surface of the chip, thermalize, and
escape from the trap.

As shown in the previous section the four-beam trap allows atoms
to be confined and well separated from the source light beams. The
translation of the trapped atoms along the x- and y-axis of this
trap preserves that separation of the central interference fringe
from the source beams. Fig.\,7 shows the intensity distribution of
the light field in the X0Z plane while the phases of the beams 3
and 4 of the four-beam trap are changed by the same amount
relative to the beams 1 and 2 (see Fig.\,4). The parameters of
this four-beam trap are the same as ones used in the previous
section. One can see that the centre of the trap is smoothly
moving along the x-axis while the phase of the beams 3 and 4 is
changed. During the shift of the central fringe the second
interference fringe from the side opposite to the direction of
motion of the central fringe is developed. Nevertheless these two
fringes stay well separated from each other by a distance of about
3\,$\mu$m. By choosing the proper intensities of the laser beams
and the rate of the spatial translation of the fringe it is
possible to keep the tunnelling probability of the translated
atoms to the other interference fringe negligibly small. The
procedure described provides a unidirectional shift of the trapped
atoms along the x-axis. Fig.\,8 shows the evolution of the
intensity distribution of the trap along the x-axis taken at
z=8.885\,$\lambda$, where the maximum of the fringe is located, as
a function of the phase difference between the two pairs of the
light beams. The dashed lines show the borders of the trap, which
correspond to the x-coordinates of the centres of the four
trapping light beams. From the last graph of Fig.\,8 it follows
that at $\phi=\pi$ the position of the shifted interference fringe
reaches the border of the trap. Let us suppose now that there are
two four-beam traps are set up next to each other in such a way
that they have a common border with two pairs of their light beams
are superimposed as shown in fig.\,9. The adiabatic transfer of
atoms between these two traps can be organized as follows. First
the second trap is completely off. By variation of the phase
difference between the light beams of the first trap form 0 to
$\pi$ atoms are moved towards the common border. Then the
intensity of the first trap is switched off while the intensity of
the second trap is simultaneously switched on. The phases of the
beams in the second trap are set in such a way to pick up the
atoms at the common border of the traps. Finally the adiabatic
change of the phases of the light beams of the second trap moves
atoms to its centre and the process of transfer of atoms between
the two traps is completed. This elementary process can be
expanded to a large number of neighboring traps so that an
all-light conveyor belt similar to the magnetic one \cite{Hansel}
can be realized.

The next important type of atomic manipulation is the coherent splitting of the atomic ensemble into two identical parts.
Such a coherent beam splitter is a key element of any interferometer. In the last few years the interferometers based on
conservative traps, which can be adiabatically transformed into double-well traps, were actively investigated \cite{Hansel_2,
Shin_1, Shin_2, Schumm}. In \cite{Hansel_2} it is proposed to split the atomic ensemble by adiabatic
transportation of a single well potential into a double well trap and then to recombine the two wells back into a single well trap.
Some possible phase
difference between the atoms in two "arms" of the interferometer would lead to variation of the population of the excited oscillation
states of that final single well trap. A simpler approach is used in \cite{Shin_1, Shin_2, Schumm}, where it is proposed to split atoms coherently
in a double well trap and then to release atoms by switching off the trap potential. As a result interference fringes like
a classical Young's double slit interferometer were observed. The experiments \cite{Shin_2, Schumm} relied on magnetic
microtraps. A very different approach was demonstrated in \cite{Shin_1} where an optical double well trap was used. This optical
trap consists of two Gaussian beam traps of different frequencies, which do not interfere with each other in a sense that they
do not produce a stationary interference pattern. A smooth spatial translation of these traps with respect to each other allows
both merger into one trap and separation to form a double well potential.

The splitter based on the interference dipole traps can also be made by placing two non-interfering traps next
to each other. The difference is that the transformation of the dipole potential is achieved now, not by moving of the
trapping light beams, but by changing their relative phases. The corresponding configuration of the trap,
which consists of six co-propagating laser beams is shown in Fig.\,9. This figure shows the intensity distribution of
light in the X0Y plane, where the waists of all the beams are located. The distance between the adjacent beams along the x-
and y-axes is d=3\,$\lambda$ and their waist radii are $w_0=\lambda$. The amplitude of the light field in the beams
3 and 4 is by factor $\sqrt{2}$ larger compared to the beams 1, 2, 5 and 6. The arrows shows the direction of the
polarization of light in each of the beams. This configuration corresponds to the two four-beam traps with crossed
polarizations, which have a common side located at x=0. We suppose that initially the phases of the outer beams 1, 2, 5 and
6 $\phi_1=\phi_2=\phi_5=\phi_6=\pi$ are in anti-phase with the phases of the central beams 3 and 4 $\phi_3=\phi_4=0$.
In these conditions the interference of the beams forms a fringe, the maximum of which is located at x=y=0, z=8.43\,$\lambda$.
The intensity distribution of light in that fringe along the x-axis taken at z=8.43\,$\lambda$ is shown by the highest curve
in fig.\,10. The other curves in this figure shows the stages of the transformation of this single well potential into a
double well trap while the phases of the outer beams are changed gradually from $\pi$ to 0.
Finally at $\phi_1=\phi_2=\phi_5=\phi_6=0$ the two maximums of the double well trap are located exactly at the centres
of the two adjacent four-beam traps at x=$\pm 1.5 \, \lambda$.

It is possible to perform a coherent splitting of atoms between the two four-beam traps formed by six beams of the
same polarization by changing their relative phases. In that case an additional interference between the outer most
beams of the traps leads to more complicated evolution of the trap potential.

\section{Talbot optical lattices}

The idea of trapping atoms in the dipole potential of a standing light wave has been put forward in 1968 \cite{Letokhov} long
before the laser cooling of atoms was achieved. The potential formed by a standing light wave is unique,
because it has an ideally sinusoidal form of periodicity, determined by the wavelength of light.
Standing light waves were used in a huge number of experiments on diffraction and interference of thermal atomic beams
\cite{Berman, Metcalf}.
A new interest to the standing light waves appeared after Bose-Einstein condensates of cold atoms
became available. The BEC loaded into three-dimensional lattices has allowed one for the first time to get filling factor
more than 1 and to observe such an effect as a superfluid - Mott insulator transition \cite{Bloch}.
The interest of the standing light waves might be explained also by the fact that an essential part of proposals for
quantum computing rely on optical lattices \cite{QC}. The advantage of the optical lattices in the quantum
computing applications is the the possibility of performing the same quantum logic operations in parallel over a large
number of identical sites of the lattice.
However there are still some unsolved problems with addressing and selective manipulation of atoms between
several distinct sites of optical lattices based on standing light waves.

In \cite{Ertmer_1, Ertmer_2} a two-dimensional lattice of about 80 dipole traps, separated from each other
by a macroscopic distance of 125\,$\mu$m has been demonstrated. The lattice was made by focussing a laser beam with a
microfabricated array of microlenses. Addressing of a single site of the lattice and modification of the
trapping potential with an additional inclined laser beam was demonstrated experimentally.

An alternative approach is to trap of atoms using fractional Talbot optical lattices, which are based on the
interference of an array of multiple co-propagating equidistant laser beams. Compared to a case of an array of
independent dipole traps \cite{Ertmer_1, Ertmer_2}, the Talbot optical lattice permits a much
smaller period of the lattice and control of the form of all or part of the interference potential by
changing the phases and amplitudes of the individual source beams.

The Talbot effect \cite{Talbot} consists of reproducing the transverse spatial distribution of light coming through a
transmission diffraction grating at the distances from the grating, which are multiples of a Talbot period
$T=2 d^2/\lambda$, where $d$ is a period of the diffraction grating. For a grating with infinite number of slits,
the transverse spatial distribution of light intensity at Talbot distances is exactly the same as the intensity distribution
of light in the plane of the grating. This is valid for both  a one-dimension and two-dimension gratings of a
definite period.

At fractional distances $T \times M/N$, where $M$ and $N$ are
coprime integers, the fractional Talbot images are observed
\cite{FT}. The transverse period of the intensity distribution of
light at fractional Talbot distances is $2 d/N$ for even $N$ and
$d/N$ for odd values of $N$. The fractional Talbot optical
images of different orders are separated from each other in
space, because they are located at different distances from the
grating and also shifted with respect to each other
transversally.

These periodic spatial distributions of light formed at fractional Talbot distances by the interference of a periodic
array of light beams can be used as atom traps.

We will skip the case of a one-dimensional Talbot grating and describe the most interesting case of the square two-dimension
grating, which is formed by 5\,$\times$ \,5 square array of 25 co-propagating Gaussian laser beams. The waists of the beams
are located in the X0Y plane with coordinates of their centres $x_j=n d$ and $y_j=m d$, where $n,m=-2,\,-1,\,0,\,1,\,2$ and
$d=10\,\lambda$ is the period of the array.  The waist radiuses of each of the beams are $w_0=\lambda$.
The corresponding Talbot period of the lattice is $T=200\,\lambda$.
Fig.\,11\,a shows the spatial intensity distribution of the corresponding interference pattern in the XZ plane at y=$d/2$.
That specific plane was taken to see the intensity maximums of the fractional Talbot lattice at $z=T/2=100\,\lambda$,
where the interference fringes are shifted by half of a period with respect to the position of the source beams
in the X0Y plane.
In full accordance with theory the transverse period of the $T/2$-lattice is equal to $d$.
The other fractional Talbot lattice of nearly same intensity and of twice smaller period $d/2$ is located at $z=T/4=50\, \lambda$.
Fig\,11\,b and 11\,c show the transverse distributions of the light intensity in the XY plane, taken at $z=T/2$
and $z=T/4$ correspondingly.
To get a better idea on the intensities of the different fractional Talbot lattices, the distribution of the
light intensity along the $z$-axis at $x=y=d/2$ is presented in fig.\,12\,a.
One can see that the intensity of the $T/4$-lattice is comparable to the intensity of the $T/2$-lattice.
Calculations show that the intensity of the interference fringes at the full Talbot distance $z=T=200\,\lambda$, the
locations of which in the XY plane are the same as the positions of the source beams, is about four times smaller compared
to the fringes of the $T/2$-lattice.
Finally, fig.\,12\,b shows the transverse distribution of the light intensity in the $T/2$-lattice taken along the
x-axis at $z=T/2$ and $y=d/2$.

To estimate the potential depth of the Talbot lattices we use again an example of Rb$^{87}$ atoms and the wavelength of
light of a Nd:YAG laser. For the laser power of 4\,mW in each of the 25 source beams of the lattice the potential depth
of the central sites of the $T/2$-lattice is about 91\,$\mu$K. The oscillation frequencies of the atoms near the
centres of the sites of the lattice are $f_z=1.74$\,kHz and $f_x=f_y=16.25$\,kHz.

Due to the large number of the source beams involved in the Talbot lattices, they offer a huge variety of ways to
manipulate the trapped atoms by changing the phases and amplitudes of different groups of the source beams.
The methods described in the previous section for the four-beam interference
traps can be easily extended to the Talbot lattices. For example, a change of the relative phases and amplitudes of
different rows of the source beams can provide a unidirectional transfer of atoms between the rows of the Talbot lattice.
The shift of the phases of the even source beams with respect to the odd ones can change the period of the Talbot lattice
potential by factor of two \cite{Ovchinnikov_3}.

Another important difference of the Talbot lattice compared to the standing light wave lattices is that the form of its
potential can essentially differ from a sinusoidal form. The general rule is that the intensity distribution of light in the
Talbot lattice reproduces the corresponding distribution of the light beams in the source plane X0Y. Therefore, the
different sites of the Talbot lattice can be well isolated from each other, which is important for applications
where the tunnelling of atoms between different sites of the lattice is undesirable.

\section{Outlook}

In this paper the interference traps formed by co-propagating Gaussian laser beams have been considered.
The same principles are also valid for the light beams formed by diffraction of light at
round holes. The construction of an all-light atom chip might be as follows.
The desired matrix of holes is made in a non-transparent screen, which is mounted on top of some optical substrate.
Micro-electro-optical elements are positioned right above or below each of the holes to control
the amplitudes and phases of the individual source light beams. This construction would allow one to address each of the
beams with electrical signals. It was demonstrated that interference traps are scalable, which means all sizes of
the traps and their distance from the waists of the source beams depend on the sizes of the beams and the distances between
them. Therefore using holes of different diameter in the chip would allow manipulation of
the spatial position of atoms not only along but also across the surface of the chip.

An alternative chip construction might use an array of single mode polarization mantaining fibres, which
could each be addressed individually.

The negative sign of the frequency detuning of the trapping light of the interference traps
and Talbot lattices provides an opportunity to use these traps for
trapping such atoms as Sr, Yb, Ca and some other atoms in a regime
when the optical Stark shift of the intercombination transition of
these atoms is cancelled \cite{Katori}. Thus fractional Talbot
lattices may provide an alternative option for the construction of
portable optical frequency standards.

\section{Conclusion}

The dipole traps based on the interference between multiple co-propagating laser beams can be configured to
enable atom manipulation by changing the phases, polarizations and intensities of the individual light beams.
Such basic manipulation of cold atoms in these traps as their spatial displacement and coherent splitting in
two identical parts have been outlined. A Talbot optical lattice dipole trap based on the interference of arrays of
co-propagating laser beams is proposed. It is shown that single sites of the Talbot lattice can be well isolated
from each other. In addition the Talbot lattice provides a number of ways to transfer atoms between lattice sites.
The variable period of the considered Talbot optical lattices enables one to choose the right compromise between the
addressability of the individual sites of the lattice and the interaction of the sites with each other.

It is shown that with rather moderate power (1-4\,mW) laser beams with wavelength 1\,$\mu$m, the depth of
interference traps for Rb$^{87}$ atoms can be $\sim$100\,$\mu$K, while the coherence time is of the order of 1 second.

This technique offers the opportunity of building a new generation
of all-light atom chips based on interference traps, which might
find applications in the design of integrated elements of quantum
logic, atom interferometers and other atom optics components.

The examples of the interference traps and their manipulation presented in this paper do not pretend to be the most optimal
ones, but are intended only to demonstrate the main principles. The most optimal and efficient schemes for interference traps
and methods of atom manipulation still have to be developed.

\section{Acknowledgments}

Many thanks to Hugh Klein for his valuable comments to the paper.
This work was funded by the UK National Measurement System Directorate of the Department of Trade and Industry
via the Strategic Research Programme of NPL.

\newpage

Fig.1 Schematics of the two-beam interference trap.

Fig.2 Spatial intensity distribution of light in the XOZ-plane of a two-beam interference trap.

Fig.3 Cross sections of the intensity distribution of light in the central interference fringe of the two-beam
interference trap, which are taken along four different directions.

Fig.4 Intensity distribution in the four-beam trap in the XOY-plane, where the waists of the beams are located.

Fig.5 Spatial intensity distribution of light in the XOZ-plane of the four-beam interference trap.

Fig.6 Cross sections of the intensity distribution of light in the central interference fringe of the four-beam
interference trap, which are taken along four different directions.

Fig.7 Intensity distribution in the four-beam trap in the
XOZ-plane as a function of the phase difference between the two
pairs of source laser beams.

Fig.8 Cross sections of the intensity distribution of light in the moving central interference fringe of the four-beam
interference trap, which are taken along the x-axis at $z=8.885\,\lambda$ for several different values of the phase shift
between the two pairs of source light beams.

Fig.9 Intensity distribution in the two superimposed four-beam
traps in the XOY-plane, where the waists of the beams are located.
The arrows shows the polarizations of the six light beams.

Fig.10 Evolution of the intensity distribution of light along the x-axis in a superposition of two four-beam interference traps,
which is taken at $z=8.43\, \lambda$ for five different values of the phases of the outer laser beams 1, 2, 5 and 6.

Fig.11 a) Intensity distribution of light in the XZ-plane at y=d/2 produced by 5\,x\,5 two-dimensional array of Gaussian light beams.
b) Intensity distribution across the fractional Talbot optical lattice taken at $z=T/2=100\, \lambda$.
c) Intensity distribution in a higher order lattice at $z=T/4=50\, \lambda$.

Fig.12 a) Cross section of the intensity distribution in the fractional Talbot lattices along the z-axis at x=y=d/2.
b) Cross section of the transverse intensity distribution along the x-axis at $z=T/2=100\, \lambda$ and y=d/2.

 \begin{figure}
 \includegraphics[scale=0.8]{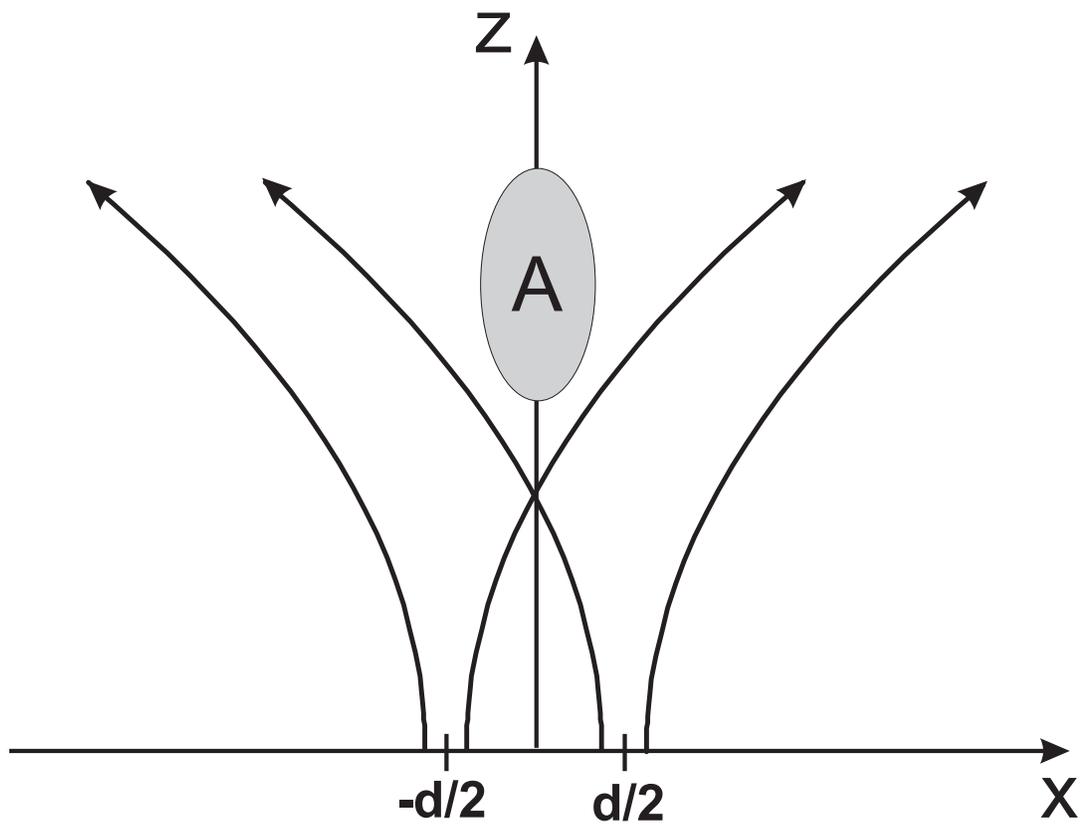}
 \caption{\label{}Yuri B. Ovchinnikov "Coherent manipulation of atoms..."}
 \end{figure}

\begin{figure}
 \includegraphics[scale=0.8]{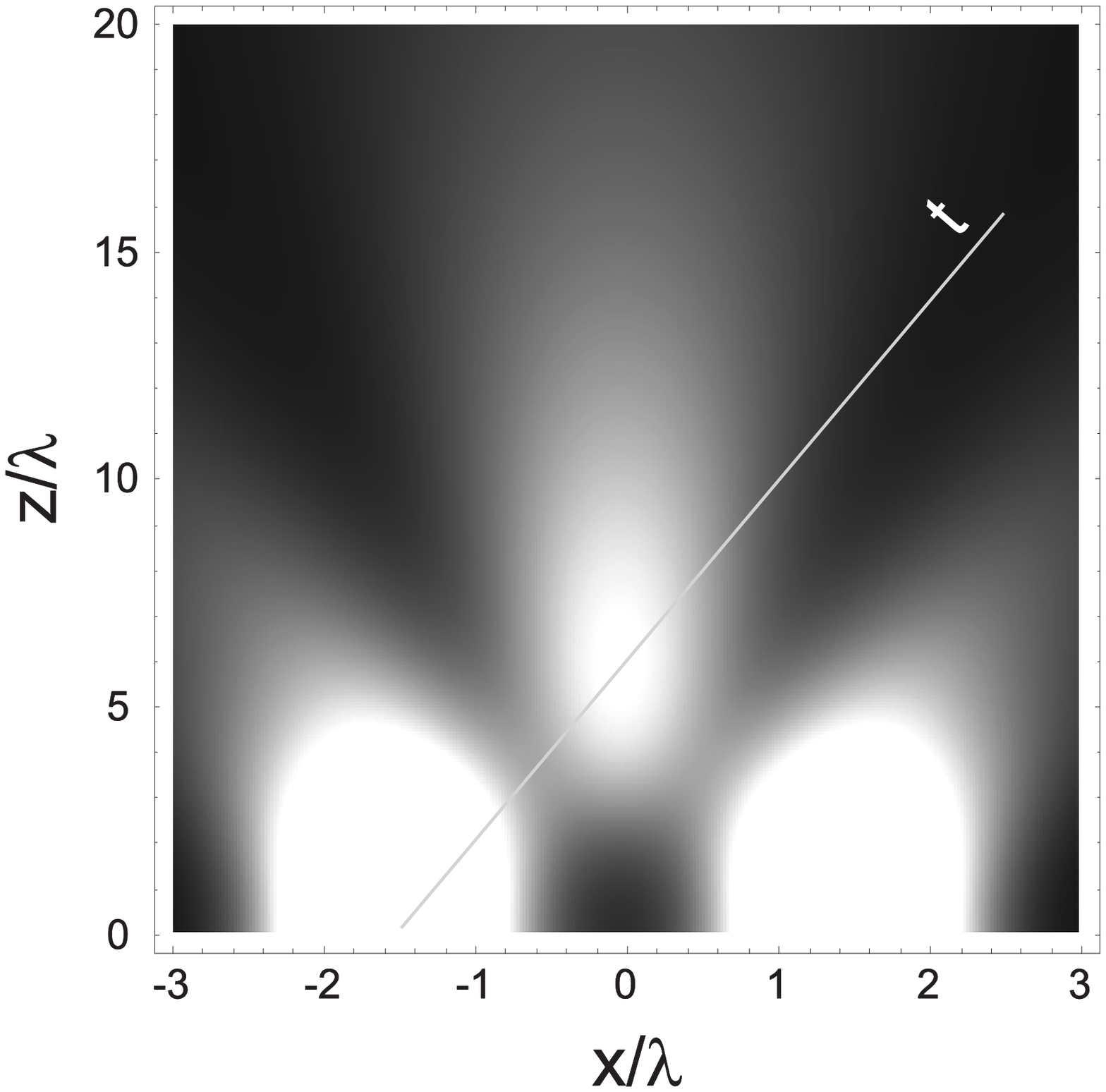}
 \caption{\label{}Yuri B. Ovchinnikov "Coherent manipulation of atoms..."}
 \end{figure}

\begin{figure}
 \includegraphics[scale=0.8]{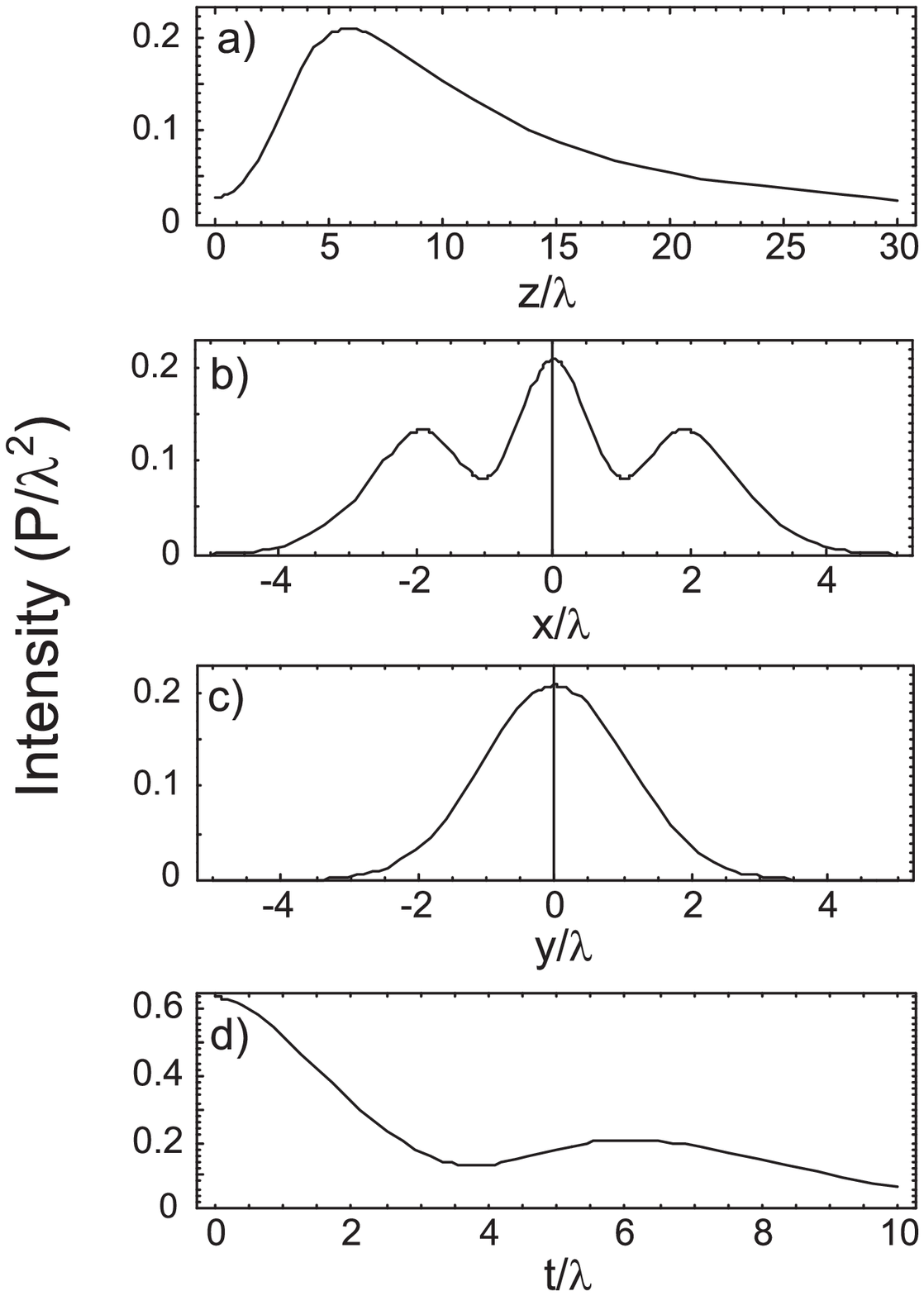}
 \caption{\label{}Yuri B. Ovchinnikov "Coherent manipulation of atoms..."}
 \end{figure}

\begin{figure}
 \includegraphics[scale=0.8]{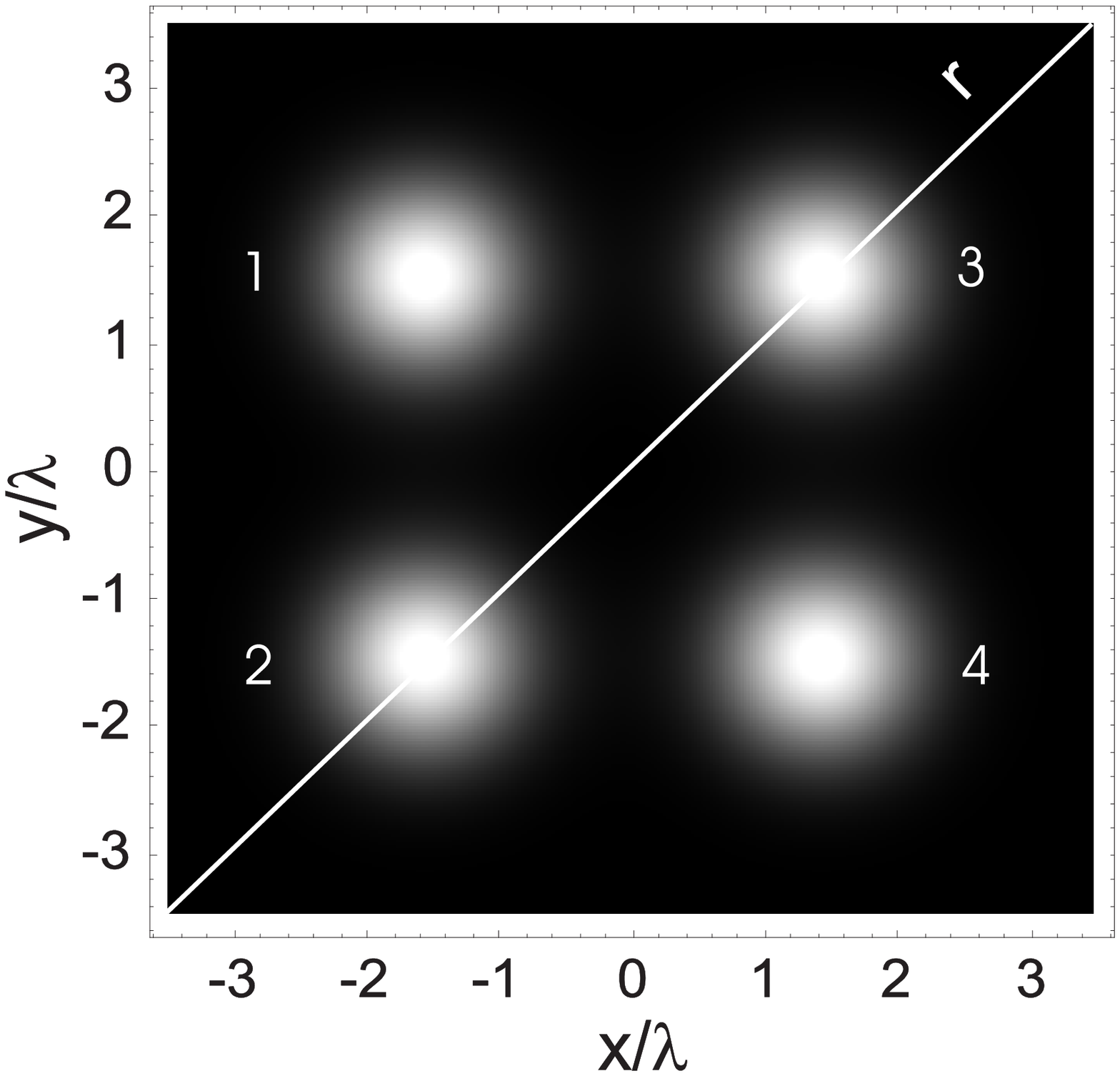}
 \caption{\label{}Yuri B. Ovchinnikov "Coherent manipulation of atoms..."}
 \end{figure}

\begin{figure}
 \includegraphics[scale=0.8]{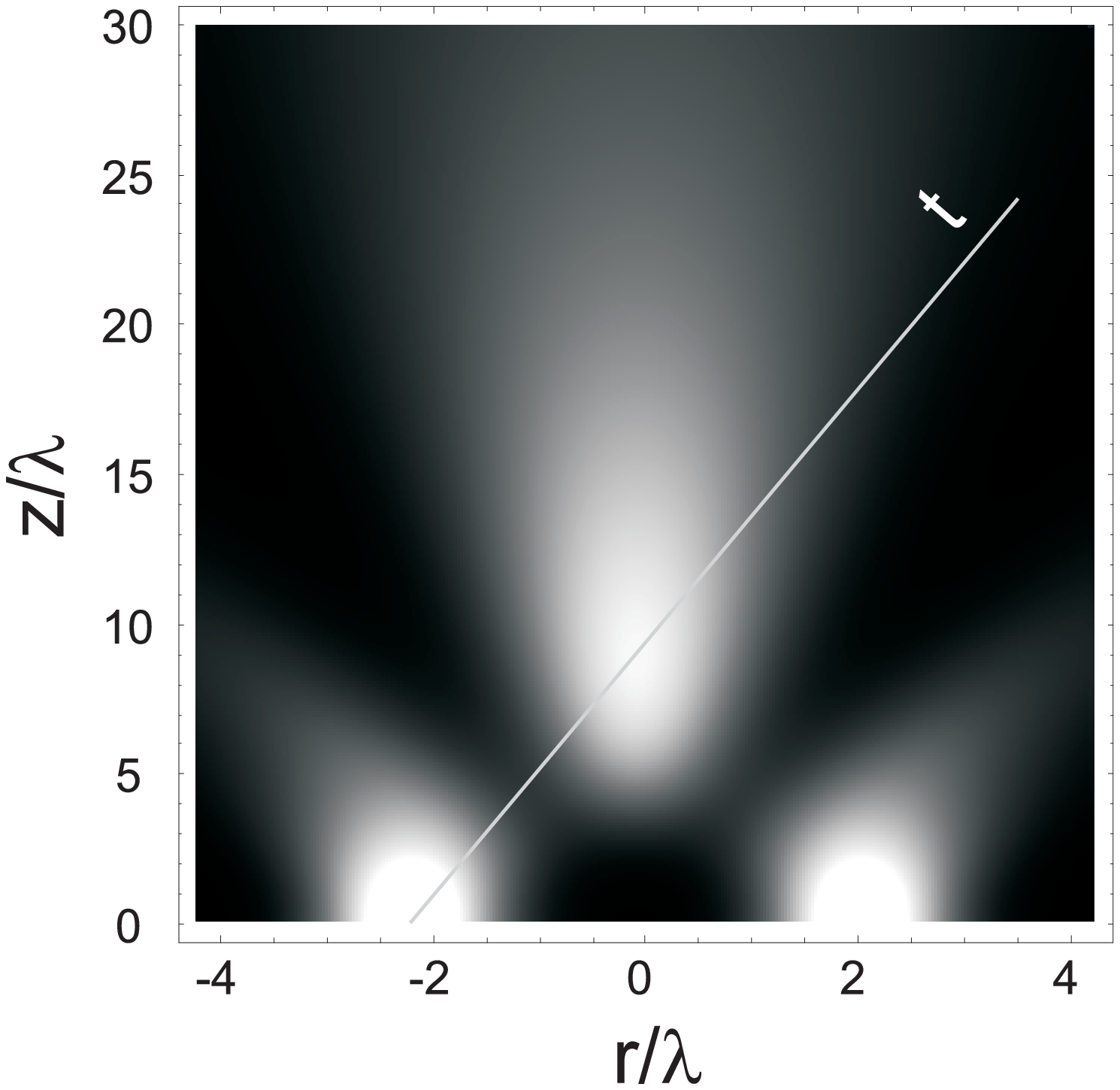}
 \caption{\label{}Yuri B. Ovchinnikov "Coherent manipulation of atoms..."}
 \end{figure}

\begin{figure}
 \includegraphics[scale=0.8]{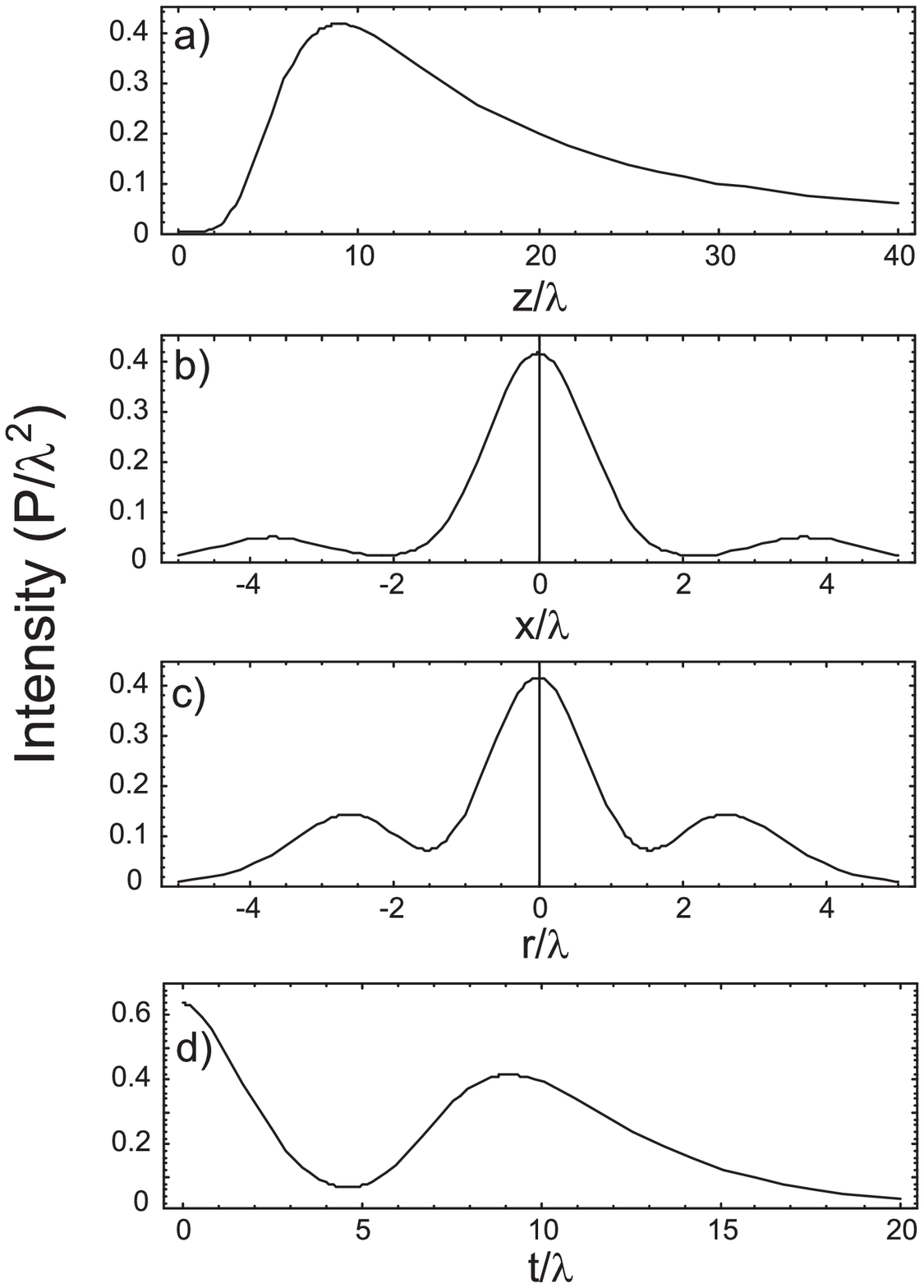}
 \caption{\label{}Yuri B. Ovchinnikov "Coherent manipulation of atoms..."}
 \end{figure}

\begin{figure}
 \includegraphics[scale=0.8]{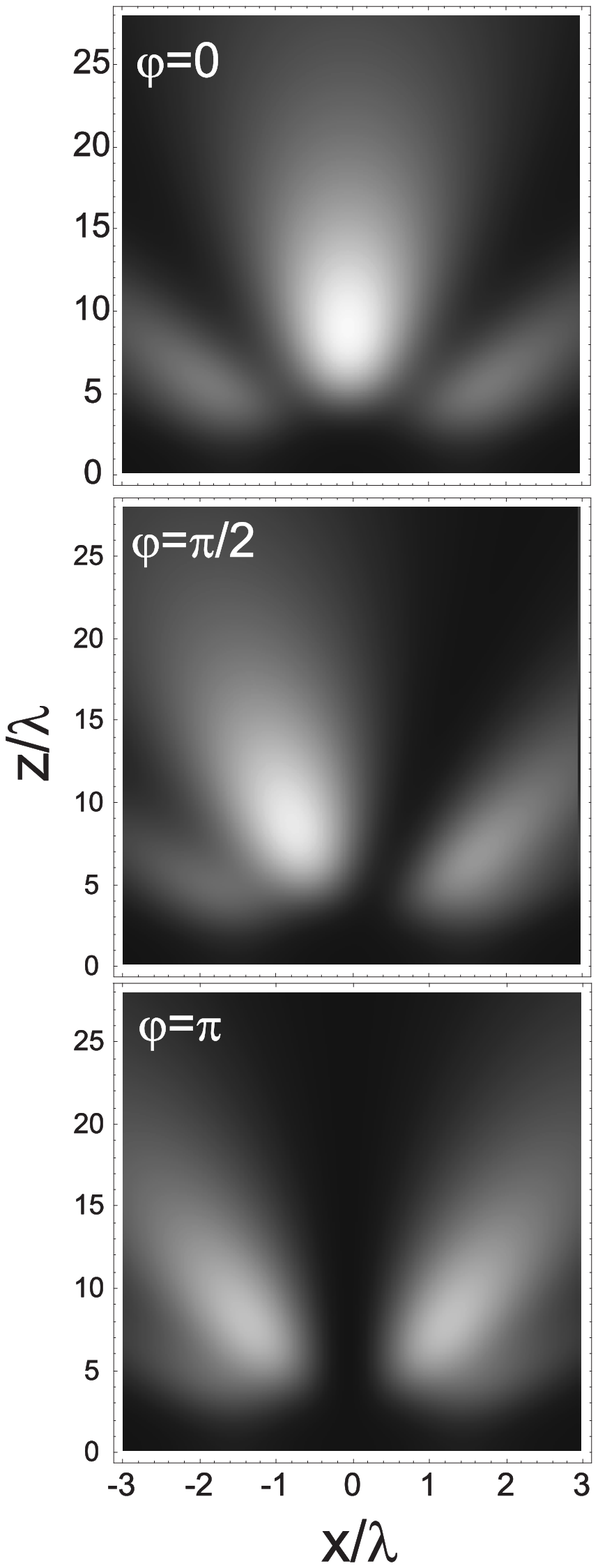}
 \caption{\label{}Yuri B. Ovchinnikov "Coherent manipulation of atoms..."}
 \end{figure}

\begin{figure}
 \includegraphics[scale=0.8]{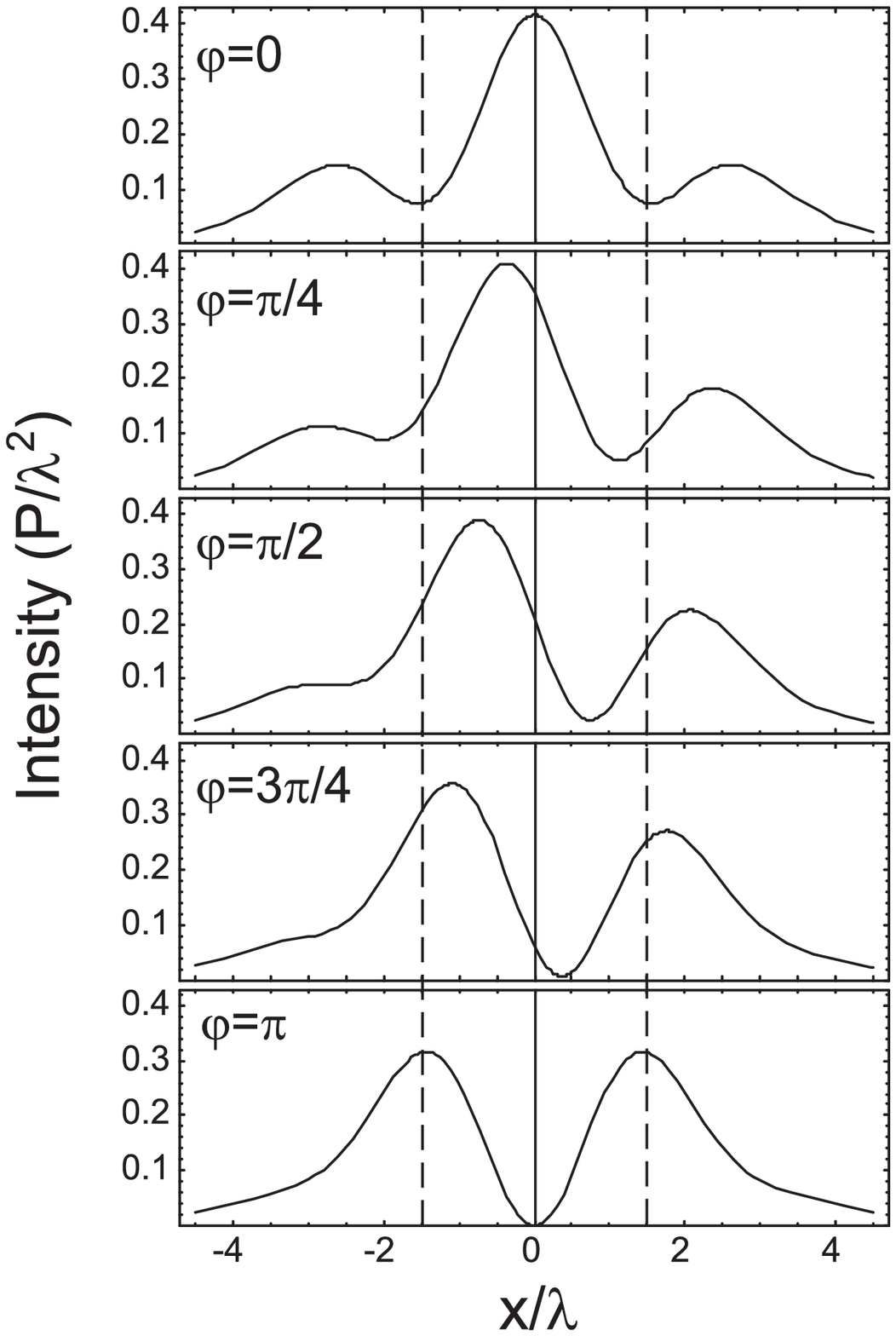}
 \caption{\label{}Yuri B. Ovchinnikov "Coherent manipulation of atoms..."}
 \end{figure}

\begin{figure}
 \includegraphics[scale=0.8]{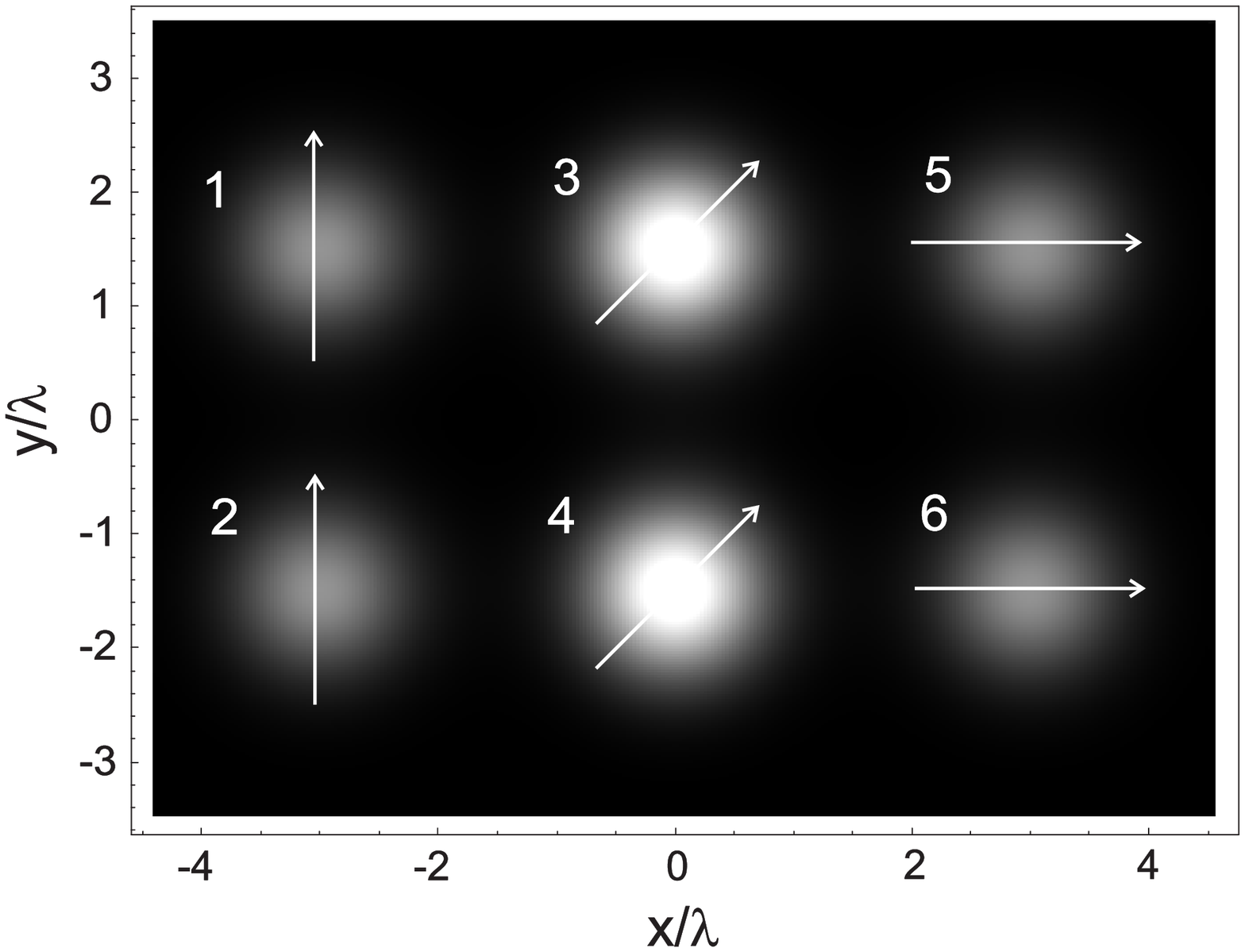}
 \caption{\label{}Yuri B. Ovchinnikov "Coherent manipulation of atoms..."}
 \end{figure}

\begin{figure}
 \includegraphics[scale=0.8]{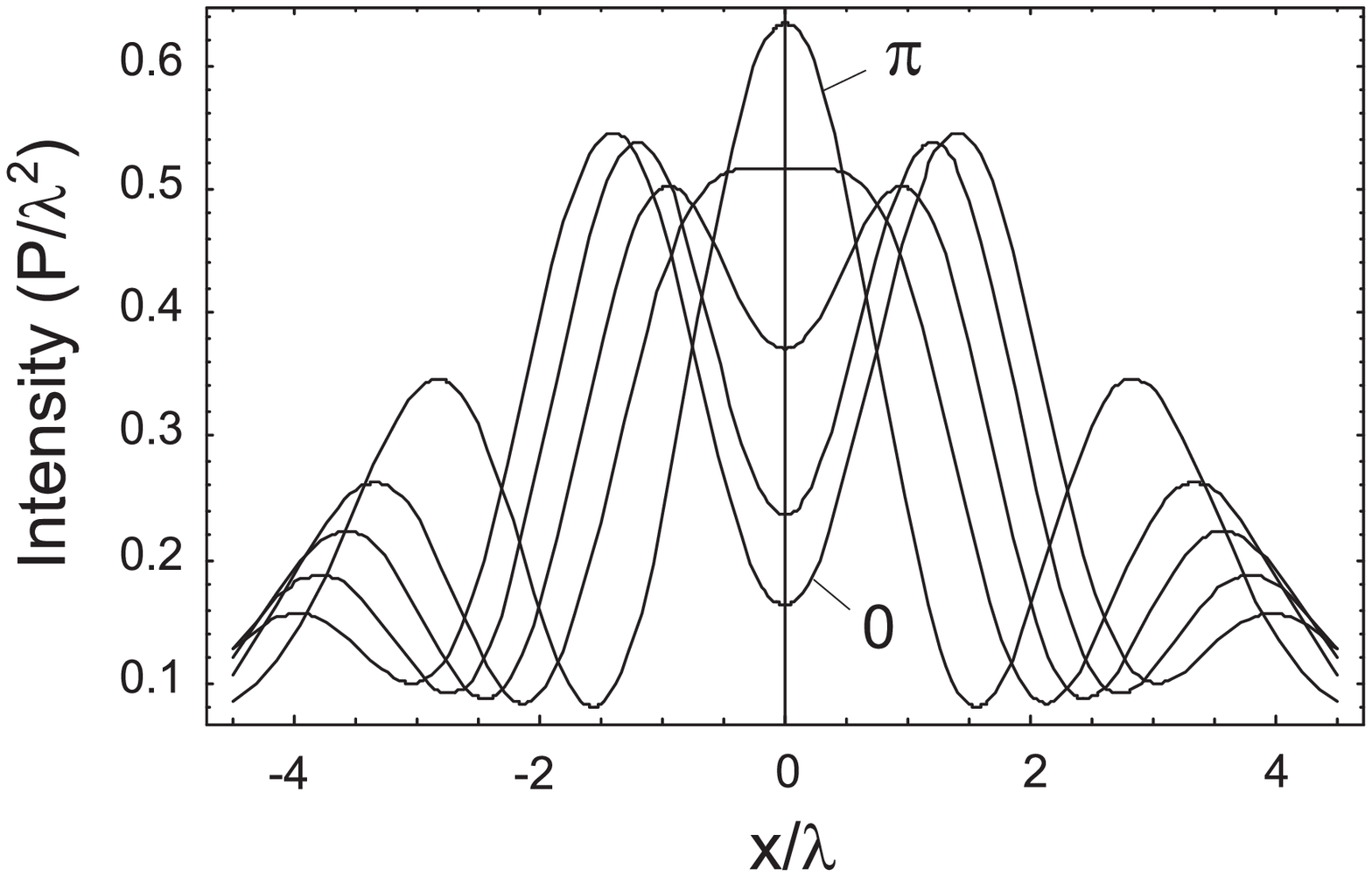}
 \caption{\label{}Yuri B. Ovchinnikov "Coherent manipulation of atoms..."}
 \end{figure}

\begin{figure}
 \includegraphics[scale=0.8]{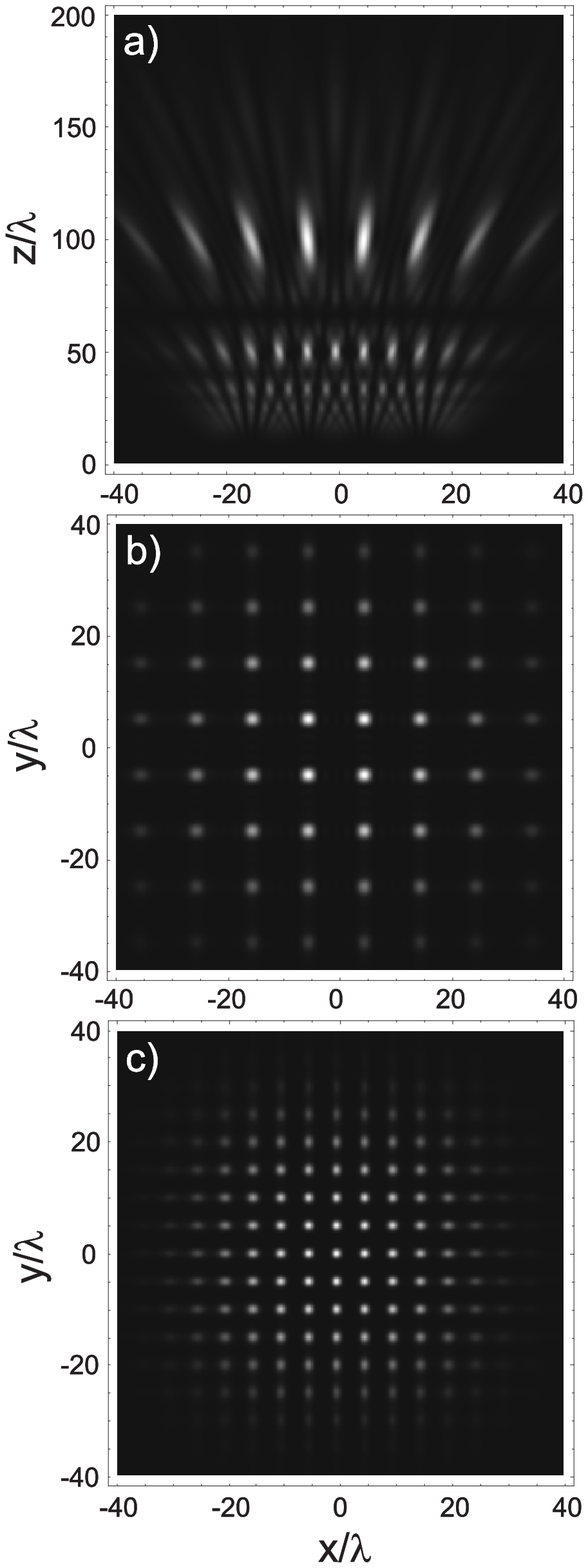}
 \caption{\label{}Yuri B. Ovchinnikov "Coherent manipulation of atoms..."}
 \end{figure}

\begin{figure}
 \includegraphics[scale=0.8]{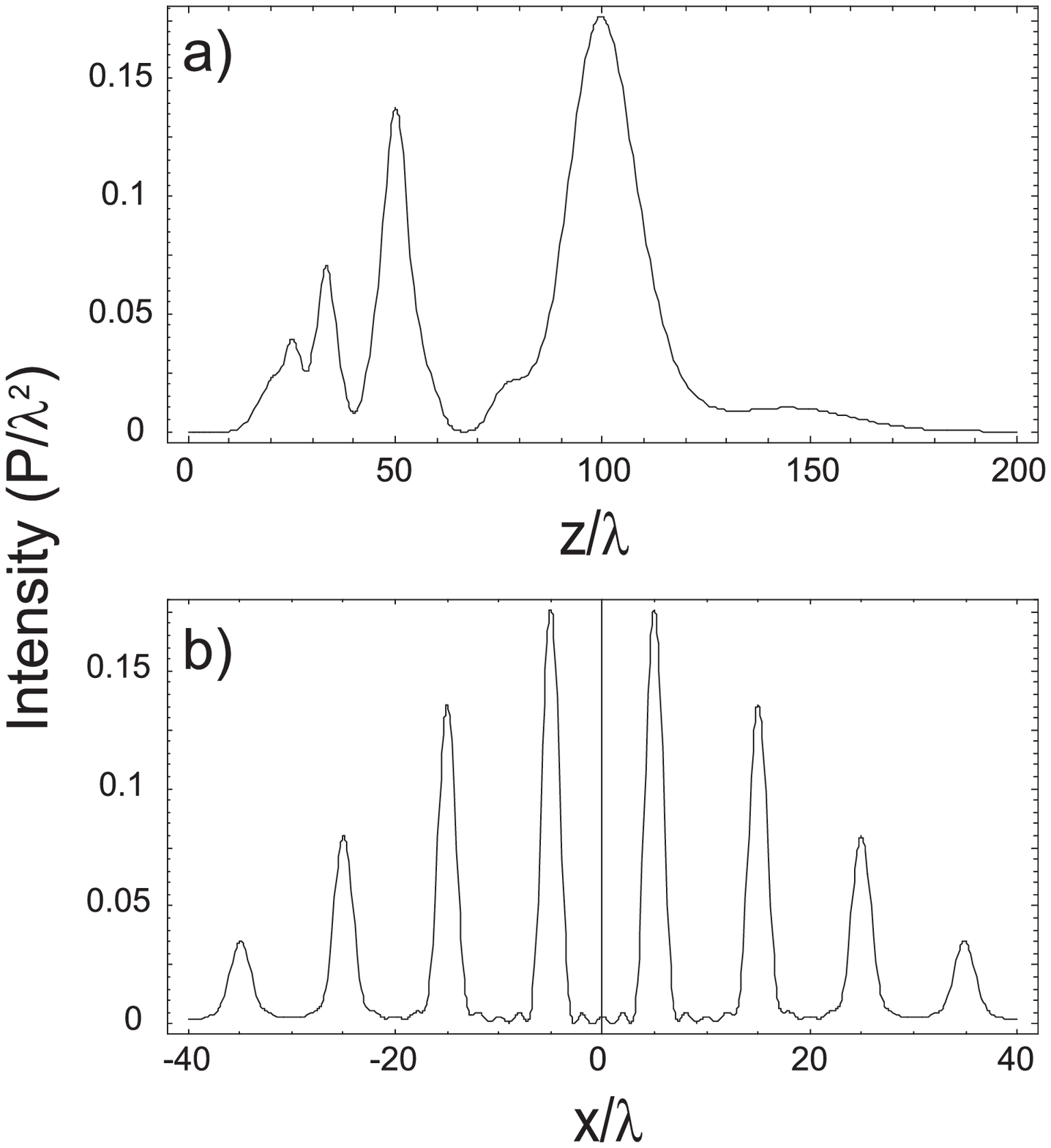}
 \caption{\label{}Yuri B. Ovchinnikov "Coherent manipulation of atoms..."}
 \end{figure}

\end{document}